%% file: main.tex
\documentclass[a4paper, amsfonts, amssymb, amsmath, reprint, showkeys, nofootinbib, twoside]{revtex4-1}
\usepackage[english]{babel}
\usepackage[utf8]{inputenc}
\usepackage[colorinlistoftodos, color=green!40, prependcaption]{todonotes}
\input{preamble}

\usepackage{amssymb}
\usepackage{natbib}

\usepackage[LGRgreek]{mathastext} 

\usepackage{lineno}
\usepackage{setspace}
 \usepackage{hyperref}

\begin{document}
\title{MoS\textsubscript{2} nanosheets incorporated $\alpha$-Fe\textsubscript{2}O\textsubscript{3}/ZnO nanocomposite with enhanced photocatalytic dye degradation and hydrogen production ability}

\author{Angkita Mistry Tama \textit{$^{a,b}$}, Subrata Das \textit{$^{a,b}$}, Sagar Dutta\textit{$^{a}$}, M D I Bhuyan\textit{$^{a}$}, \\ M. N. Islam\textit{$^{c}$} and M. A. Basith\textit{$^{a}$}}
    \email[Email address: ]{mabasith@phy.buet.ac.bd}
    \affiliation{\\ \textit{$^{a}$}Nanotechnology Research Laboratory, Department of Physics, Bangladesh University of Engineering and Technology, Dhaka-1000, Bangladesh.\\
    \textit{$^{c}$} Department of Chemistry, Bangladesh University of Engineering and Technology, Dhaka-1000, Bangladesh.\\
    \textit{$^{b}$} These authors contributed equally.
    \\ \\ DOI: \href{https://doi.org/10.1039/C9RA07526G}{10.1039/C9RA07526G}}


\begin{abstract}
We have synthesized MoS\textsubscript{2} incorporated $\alpha$-Fe\textsubscript{2}O\textsubscript{3}/ZnO nanocomposites by adapting a facile hydrothermal synthesis process. The effect of incorporating  ultrasonically exfoliated few-layer MoS\textsubscript{2} nanosheets on the solar-light driven photocatalytic performance of $\alpha$-Fe\textsubscript{2}O\textsubscript{3}/ZnO photocatalyst nanocomposites has been demonstrated. Structural, morphological and optical characteristics of the as-synthesized nanomaterials are comprehensively investigated and analyzed by performing Rietveld refinement of powder X-ray diffraction patterns, field emission scanning electron microscopy and UV-visible spectroscopy, respectively. The photoluminescence spectra of the as-prepared nanocomposites elucidate that the recombination of photogenerated electron-hole pairs is highly suppressed due to incorporation of MoS\textsubscript{2} nanosheets. Notably, the ultrasonicated MoS\textsubscript{2} incorporated $\alpha$-Fe\textsubscript{2}O\textsubscript{3}/ZnO nanocomposite manifests 91\% and 83\% efficiency in degradation of rhodamine B dye and antibiotic ciprofloxacin respectively under solar illumination. Active species trapping experiments reveal that the hydroxyl ($\cdot OH $) radicals play a significant role in RhB degradation. Likewise the dye degradation efficiency, the amount of hydrogen produced by this nanocomposite via photocatalytic water splitting is also considerably higher as compared to both non-ultrasonicated
MoS\textsubscript{2} incorporated $\alpha$-Fe\textsubscript{2}O\textsubscript{3}/ZnO and $\alpha$-Fe\textsubscript{2}O\textsubscript{3}/ZnO nanocomposites as well as Degussa P25 titania nanoparticles. This indicates the promising potential of the incorporation of ultrasonicated MoS\textsubscript{2} with $\alpha$-Fe\textsubscript{2}O\textsubscript{3}/ZnO nanocomposite for generation of carbon-free hydrogen by water splitting. The substantial increase in the photocatalytic efficiency of $\alpha$-Fe\textsubscript{2}O\textsubscript{3}/ZnO after incorporation of ultrasonicated MoS\textsubscript{2} can be attributed to its favorable band structure, large surface to volume ratio, effective segregation and migration of photogenerated electron-hole pairs at the interface of heterojunction and the plethora of exposed active edge sites provided by few-layer MoS\textsubscript{2} nanosheets.
\end{abstract}


\maketitle

\input{sections/mainmanuscript.tex}  

\bibliography{PCCP} 
\bibliographystyle{rsc} 


\end{document}

%% file: preamble.tex
\usepackage{amsthm}
\usepackage{mathtools}
\usepackage{physics}
\usepackage{xcolor}
\usepackage{graphicx}
\usepackage[left=23mm,right=13mm,top=35mm,columnsep=15pt]{geometry} 
\usepackage{adjustbox}
\usepackage{placeins}
\usepackage[T1]{fontenc}
\usepackage{lipsum}
\usepackage{csquotes}

%% file: sections/mainmanuscript.tex
\section{Introduction}

In the last few decades, the unprecedented growth in population and industry has tremendously elevated the necessity of finding a clean and sustainable alternative source of energy to overcome the global energy crisis and increasing environmental pollution. Hydrogen (H\textsubscript{2}) can be considered as an ideal source of energy for the future owing to its high energy capacity along with non-polluting nature.\cite{r37, r42} Notably, semiconductor photocatalysed splitting of water by solar illumination is an effective and economical route to produce H\textsubscript{2} as an inexhaustible source of carbon-free fuel for numerous applications.\cite{r5, r23} Since 1972, after the first successful use of n-type titania (TiO\textsubscript2) photoelectrode in solar H\textsubscript2 production via water splitting,\cite{r12} metal oxide semiconductors have been extensively investigated and employed both in solar H\textsubscript{2} generation and photocatalytic decontamination and removal of toxic dyes from wastewater owing to their satisfactory photocatalytic efficiency and good stability under the electrochemical reaction conditions.\cite{r1, r3} Moreover, in recent years, semiconductor photocatalysts are being employed to mineralize and eliminate second and third generation antibiotics such as tetracycline chloride, ciprofloxacin and their by-products from wastewater. \cite{tc1, tc2, cipro} Especially, nanoparticles of metal oxide semiconductors have achieved significant research interest in the last few decades compared to their bulk counterparts because of their high surface to volume ratio  which enables a plenty of photons to be incident upon the surface.\cite{r1, r22} In the past decades, titanium dioxide (TiO\textsubscript2) (bandgap $\sim3.2$ eV) has been studied the most as a semiconductor photocatalyst because of its low production cost, high chemical stability and non-toxic attributes.\cite{r8, r26} Recently, zinc oxide (ZnO), an n-type semiconductor oxide photocatalyst has emerged as a suitable alternative to TiO\textsubscript2 since it possesses almost same bandgap energy ($\sim$ 3.37 eV) \cite{r39} but exhibits higher quantum and absorption efficiency when compared to TiO\textsubscript2. \cite{r36} Nevertheless, the fast recombination rate of photoexcited electron-hole (e-h) pairs in ZnO along with photo-corrosion degrades its photocatalytic efficiency significantly.\cite{r31} To enhance the solar energy conversion efficiency by absorbing a broader range of the solar spectrum, hematite ($\alpha$-Fe\textsubscript{2}O\textsubscript{3}), a visible light responsive photocatalyst \cite{r28} with moderate band gap ($ \sim $2.2 eV), has emerged as a promising candidate to form nanocomposite with ZnO nanoparticles. In some previous investigations,\cite{r21, r45} a photocatalyst combining the excellence of both ZnO and $\alpha$-Fe\textsubscript{2}O\textsubscript{3} has been achieved through the development of $\alpha$-Fe\textsubscript{2}O\textsubscript{3}/ZnO nanocomposite which has manifested good performance in photocatalytic decomposition of different organic dyes. However, the solar H\textsubscript{2} production potential of this nanocomposite is mostly unexplored and the extant synthesis procedures are quite robust involving multiple steps, high temperature, especial solvents and additives.\\ 
Notably, graphene-like two-dimensional (2D) layered transition metal dichalcogenides (TMDs) with enormous surface area and agile carrier transport properties can further improve the photocatalytic abilities of metal oxides having optimal optical absorption potential.\cite{r14, r25, r17, r20, r41} Molybdenum disulfide (MoS\textsubscript{2}) is a typically used TMDs which has been reported to introduce incredible improvements in optical absorption, electron mobility, surface morphology and adsorption capabilities of pristine metal oxide nanoparticles by forming heterojunctions with them.\cite{r40, r35} Especially, MoS\textsubscript{2} nanosheets can effectively catalyze the H\textsubscript{2} evolution reaction by utilizing their sulphur edge sites while keeping basal planes catalytically inert.\cite{r16} In a recent investigation,\cite{r10} we have adapted a facile ultrasound driven exfoliation technique to obtain MoS\textsubscript{2} nanosheets from bulk MoS\textsubscript{2} powder. We have observed that few-layer MoS\textsubscript{2} nanosheets exfoliated from bulk MoS\textsubscript{2} powder via such a facile one-step ultrasonication technique \cite{r4} demonstrate better photocatalytic performance than their bulk counterpart.\\
Inspired by such improvements, in this investigation, we have synthesized $\alpha$-Fe\textsubscript{2}O\textsubscript{3}/ZnO nanocomposite and afterward, incorporated it with both bulk MoS\textsubscript{2} powder and few-layer MoS\textsubscript{2} nanosheets distinctly. For brevity, non-ultrasonicated MoS\textsubscript{2} and ultrasonicated MoS\textsubscript{2} will be referred to as NMS and UMS respectively, from now on. We have adapted a step-wise low temperature hydrothermal synthesis technique \cite{r6} to obtain $\alpha$-Fe\textsubscript{2}O\textsubscript{3} nanoparticles as well as nanocomposites i.e. $\alpha$-Fe\textsubscript{2}O\textsubscript{3}/ZnO, NMS incorporated $\alpha$-Fe\textsubscript{2}O\textsubscript{3}/ZnO and UMS incorporated $\alpha$-Fe\textsubscript{2}O\textsubscript{3}/ZnO. Structural, morphological and optical characterizations of all the as-synthesized samples have been carried out and their photocatalytic performance has been compared to that of commercially available titania, TiO\textsubscript{2} (Degussa P25) nanoparticles in photodegradation of a representative organic pollutant rhodamine B (RhB) dye by solar irradiation. We have observed that UMS incorporated $\alpha$-Fe\textsubscript{2}O\textsubscript{3}/ZnO nanocomposite exhibits considerably better performance in dye degradation than $\alpha$-Fe\textsubscript{2}O\textsubscript{3} nanoparticles, $\alpha$-Fe\textsubscript{2}O\textsubscript{3}/ZnO, NMS incorporated $\alpha$-Fe\textsubscript{2}O\textsubscript{3}/ZnO nanocomposites and titania nanoparticles. This nanocomposite has also photodegraded the antibiotic ciprofloxacin in aqueous medium with high efficiency. Further, the UMS incorporated $\alpha$-Fe\textsubscript{2}O\textsubscript{3}/ZnO nanocomposite has demonstrated higher potential to generate H\textsubscript{2} via solar-driven water-splitting in comparison with other synthesized nanocomposites under scrutiny. We have also performed active species trapping experiments to evaluate the role of different reactive species in the degradation of RhB dye. Finally, a model has been proposed to critically assess the underlying mechanism behind the superior photocatalytic performance of UMS incorporated $\alpha$-Fe\textsubscript{2}O\textsubscript{3}/ZnO nanocomposite.

\section{Experimental Details}

\subsection{Materials}

 Analytical grade zinc nitrate hexahydrate [Zn(NO\textsubscript3)\textsubscript2.6H\textsubscript2O], ferric nitrate nonahydrate [Fe(NO\textsubscript3)\textsubscript3.9H\textsubscript2O], sodium hydroxide [NaOH], ammonium hydroxide [NH\textsubscript4OH] and molybdenum disulfide [MoS\textsubscript{2}] powder (Sigma-Aldrich, 99\% trace metals basis) were used as chemical reagents. 
\subsection{Sample preparation}
\subsubsection*{Synthesis of $\alpha$-Fe\textsubscript{2}O\textsubscript{3} nanoparticles.}The $\alpha$-Fe\textsubscript{2}O\textsubscript{3} nanoparticles were prepared using the hydrothermal reaction technique \cite{r6, r43} by dissolving 1 mmol of  Fe(NO\textsubscript3)\textsubscript3.9H\textsubscript2O in 50 mL deionized (DI) water. By adding NH\textsubscript4OH, the pH value of the solution was maintained at 9. After stirring rigorously for 50 minutes using a magnetic stirrer, the solution was placed in a Teflon-lined autoclave \cite{r6} and put in a programmable oven for 10 hours. The temperature of the oven was maintained at 120 $^{\circ}$C. The obtained suspension was then cooled down to room temperature (RT) followed by centrifugation using centrifuge machine (Hettich Universal 320) and consequent rinsing with DI water and ethanol and afterward, was dried at 80 $^{\circ}$C for 10 hours.\cite{r6} The synthesized $\alpha$-Fe\textsubscript{2}O\textsubscript{3} nanoparticles were used to prepare $\alpha$-Fe\textsubscript{2}O\textsubscript{3}/ZnO nanocomposite  by adapting a similar hydrothermal technique. 

\subsubsection*{Synthesis of $\alpha$-Fe\textsubscript{2}O\textsubscript{3}/ZnO nanocomposite.} To synthesize $\alpha$-Fe\textsubscript{2}O\textsubscript{3}/ZnO nanocomposite, 1 mmol of $\alpha$-Fe\textsubscript{2}O\textsubscript{3} nanoparticles and 1 mmol of Zn(NO\textsubscript3)\textsubscript2.6H\textsubscript2O were dissolved in 50 mL of 5M NaOH solution in stoichiometric proportions. After stirring properly for 4 hours, the mixture was transferred to the autoclave and heated for 2 hours at 190 $^{\circ}$C. Then the sample was cooled down to RT naturally. Afterward, centrifugation was applied and consequent rinsing was performed using DI water and ethanol. After this purification process, the solution was dried for 10 hours at 80$^{\circ}$C to get the desired $\alpha$-Fe\textsubscript{2}O\textsubscript{3}/ZnO nanocomposite.

\subsubsection*{Synthesis of few-layer MoS\textsubscript{2} nanosheets.} Ultrathin MoS\textsubscript{2} nanosheets were synthesized from bulk MoS\textsubscript{2} powder (referred earlier as NMS) using ultrasonication assisted exfoliation technique.\cite{r10} Initially, 300 mg of MoS\textsubscript{2} powder was dissolved in 60 mL of isopropanol and ultrasonicated for 90 minutes.\cite{r4, r10} The temperature in the ultrasonic bath was maintained at 50$ \pm $ 2 $^{\circ}$C by adjusting the temperature control knob. After sonication, the resultant solution was allowed to sediment for about 4 hours to remove the residue powder and micron size thick sheets and to obtain MoS\textsubscript{2} nanosheets by collecting 35\% of the mass as supernatant.  Prior to further characterization, these supernatant (ultrasonicated MoS\textsubscript{2} nanosheets, referred earlier as UMS) were dried at 80 $^{\circ}$C for 5 hours.

\subsubsection*{Synthesis of NMS incorporated $\alpha$-Fe\textsubscript{2}O\textsubscript{3}/ZnO and UMS incorporated $\alpha$-Fe\textsubscript{2}O\textsubscript{3}/ZnO nanocomposites.}Hydrothermal synthesis technique was also used for incorporating NMS and UMS individually with $\alpha$-Fe\textsubscript{2}O\textsubscript{3}/ZnO nanocomposites. To prepare NMS incorporated $\alpha$-Fe\textsubscript{2}O\textsubscript{3}/ZnO nanocomposite, 1 mmol of $\alpha$-Fe\textsubscript{2}O\textsubscript{3}/ZnO and 0.05 mmol of non-ultrasonicated MoS\textsubscript{2} powder were added into 50 mL of DI water and stirred rigorously for 4 hours. The mixture was then placed in an autoclave and heated for 14 hours at 150 $^{\circ}$C. Finally, the desired samples were collected following the aforementioned steps i.e. centrifugation, rinsing and drying for 10 hours at 80  $^{\circ}$C. Following this same technique, ultrasonicated few-layer MoS\textsubscript{2}  nanosheets i.e. UMS incorporated $\alpha$-Fe\textsubscript{2}O\textsubscript{3}/ZnO nanocomposite was synthesized.

\subsection{Characterization techniques}
The crystallographic parameters of the synthesized samples were determined by analyzing their powder X-ray diffraction (XRD) patterns obtained using a diffractometer (PANalytical Empyrean) with Cu X-ray source. The experimentally obtained powder XRD patterns was further refined by Rietveld method using the FULLPROF package. \cite{r34} The surface morphology of the samples was observed by performing field emission scanning electron microscopy (FESEM) imaging using a scanning electron microscope (XL30SFEG; Philips, Netherlands). UV-visible diffuse reflectance spectra (DRS) of the synthesized materials were recorded for wavelengths ranging from 200 nm to 800 nm at RT using a UV-vis spectrophotometer (UV- 2600, Shimadzu). \cite{r2} Steady-state photoluminescence (PL) spectroscopy of the synthesized materials was carried out at RT using Spectro Fluorophotometer (RF-6000, Shimadzu) to qualitatively evaluate their rate of photogenerated e-h recombination during photocatalysis. PL measurements were carried out with an excitation wavelength of ~371 nm for all the samples except NMS incorporated $\alpha$-Fe\textsubscript{2}O\textsubscript{3}/ZnO nanocomposite for which the applied excitation wavelength was 347.5 nm. The excitation wavelengths were chosen by studying the absorbance spectra of the respective material determined by UV-vis spectrophotometer. 

\subsection{Photocatalytic degradation and hydrogen evolution }
The photocatalytic performance of all the synthesized nanomaterials was evaluated by photodegradation of RhB \cite{r6, r33} dye in aqueous medium. A mercury-xenon (Hamamatsu L8288, 500W) lamp having irradiance value of 100 mWcm\textsuperscript{-2} was used as a solar simulator. The amount of RhB dye in the suspension, before and after illumination in presence of photocatalyst nanomaterials, was measured by determining the intensity of the absorbance peak of the mixture by UV-visible spectroscopy. The absorbance measurements were carried out for 4 hours at 1 hour interval. For testing the reusability of the photocatalyst, the photodegradation experiment was run 4 times. After each run, the residue nanomaterial was extracted from the solution and again used as photocatalyst in the next photodegradation cycle.\cite{r1} The photocatalytic performance of UMS incorporated $\alpha$-Fe\textsubscript{2}O\textsubscript{3}/ZnO in degradation of ciprofloxacin was also evaluated by following the same experimental technique. \cite{cipro}\\
A photocatalytic hydrogen production experiment was conducted using typical method as was described in details in our previous investigation.\cite{r6} For this test, 40 mg sample was dispersed into 60 ml DI water and then irradiated by the 500 W mercury-xenon lamp. 

\subsection{Active species trapping experiment}
In order to investigate the photocatalytic mechanism of RhB dye degradation, active species trapping experiments were performed with different scavengers i.e. isopropanol (1 mmol L$^{-1}$), acrylamide (1 mmol L$^{-1}$), ethylenediaminetetraacetic acid disodium salt dihydrate (EDTA-2Na, 1 mmol L$^{-1}$ ) and potassium dichromate (K$_{2}$Cr$_{2}$O$_{7}$, 1 mol L$^{-1}$) \cite{trap1, trap2}. Notably, the experimental conditions were kept same for all photocatalytic degradation tests.

\section{Results and discussions}
\subsection{Structural characterization}
\subsubsection*{Crystal structure analysis.}

\begin{figure*}
 \centering
 \includegraphics[width= 1\textwidth]{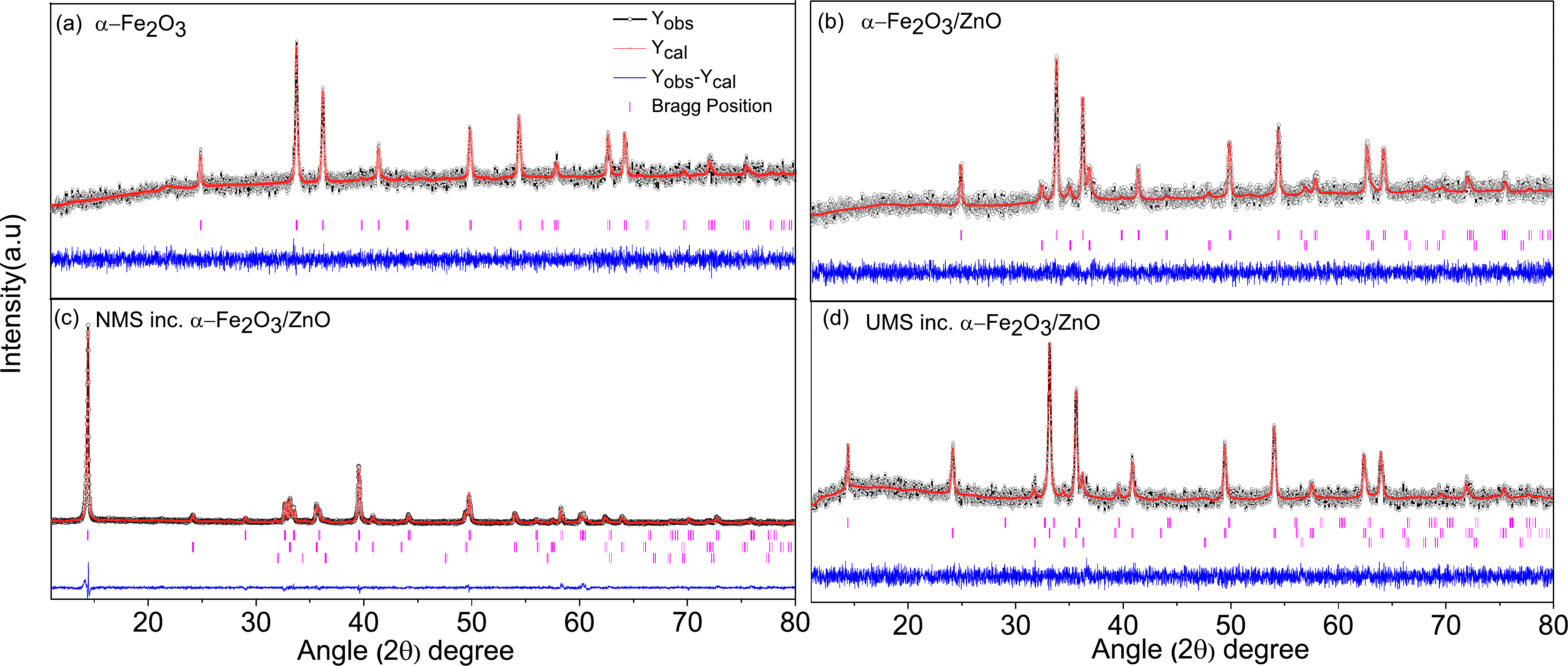}
 \caption{ Rietveld refined powder XRD patterns of (a) $\alpha$-Fe\textsubscript{2}O\textsubscript{3} nanoparticles, (b) $\alpha$-Fe\textsubscript{2}O\textsubscript{3}/ZnO, (c) non-ultrasonicated MoS\textsubscript{2} incorporated $\alpha$-Fe\textsubscript{2}O\textsubscript{3}/ZnO and (d) ultrasonicated MoS\textsubscript{2} incorporated $\alpha$-Fe\textsubscript{2}O\textsubscript{3}/ZnO nanocomposites. Due to space constraint, `incorporated' has been abbreviated to `inc.' in legends of all the figures and tables.}
\end{figure*}

Fig. 1 presents the observed and Rietveld refined powder XRD patterns of all the synthesized nanomaterials. Fig. 1(a) demonstrates the XRD pattern of the $\alpha$-Fe\textsubscript{2}O\textsubscript{3} nanoparticles. All diffraction peaks of Fe\textsubscript{2}O\textsubscript{3} can be indexed to alpha ($\alpha$)-phase having a trigonal crystal structure and R-3c space group. The XRD pattern of the $\alpha$-Fe\textsubscript{2}O\textsubscript{3}/ZnO nanocomposite is shown in Fig. 1(b) in which along with the diffraction peaks of $\alpha$-Fe\textsubscript{2}O\textsubscript{3}, some additional peaks can be observed. These peaks are matched with the peaks of hexagonal wurtzite ZnO crystal structure which belongs to P63mc space group. The obtained results are congruent with the ones reported in the literature.\cite{r11, r19, r46} Rietveld refinement of the powder XRD patterns reveals high phase purity of both $\alpha$-Fe\textsubscript{2}O\textsubscript{3} nanoparticles and $\alpha$-Fe\textsubscript{2}O\textsubscript{3}/ZnO nanocomposite since no undesired peaks were detected. The success achieved in the suppression of impurity phases might be credited to the meticulous regulation of the stoichiometry of precursor materials. 
\\The Rietveld refined XRD spectra of NMS incorporated $\alpha$-Fe\textsubscript{2}O\textsubscript{3}/ZnO and UMS incorporated $\alpha$-Fe\textsubscript{2}O\textsubscript{3}/ZnO nanocomposites are depicted in Fig. 1(c) and 1(d) respectively. Three separate phases corresponding to $\alpha$-Fe\textsubscript{2}O\textsubscript{3}, ZnO and MoS\textsubscript{2} are observed in their XRD patterns which reveals the successful formation of pristine MoS\textsubscript{2} incorporated $\alpha$-Fe\textsubscript{2}O\textsubscript{3}/ZnO nanocomposite without any impurity phases. The crystallographic phase corresponding to MoS\textsubscript{2} is found to have a hexagonal crystal structure belonging to P63/mmc space group which is consistent with the results of previous investigation.\cite{r17} Notably, as shown in Fig. 1(c), the peak introduced by MoS\textsubscript{2} at around 2$\theta$= 14.48$^{\circ}$ has significantly high intensity in the NMS incorporated sample which is substantially lower in few-layer MoS\textsubscript{2} i.e. UMS incorporated $\alpha$-Fe\textsubscript{2}O\textsubscript{3}/ZnO nanocomposite (Fig. 1(d)). Recently, in a separate investigation,\cite{r10} we have observed a noticeable decrement in the peak intensity of XRD patterns of MoS\textsubscript{2} after ultrasonication. Hence, the powder XRD patterns indicate the successful incorporation of few-layer MoS\textsubscript{2} nanosheets with $\alpha$-Fe\textsubscript{2}O\textsubscript{3}/ZnO nanocomposite.

The crystallographic phases, space groups, lattice parameters and unit cell
volume (V) for all the constituents of the synthesized nanomaterials are listed in Table 1. Noticeably, the lattice parameters and volume of the phases corresponding to $\alpha$-Fe\textsubscript{2}O\textsubscript{3} and ZnO remained almost constant in every step of the synthesis process. Hence, it can be inferred that no distortion was introduced in the crystal structure of the individual phase during synthesis. The atomic coordinates, crystallographic phases (in wt\%) of the constituent of synthesized nanomaterials and reliability (R) factors (R\textsubscript{wp}, R\textsubscript{p}, R\textsubscript{Exp}  and $\chi^{2}$) are tabulated in ESI Table S1. The diminutive values of R factors indicate excellent fit with the defined crystal structures.

\begin{table*}[t!]
\small
\caption{ Crystallographic phases, space groups, lattice parameters (a, b and c) and unit cell volume (V) for all the constituents of the synthesized nanomaterials obtained from Rietveld refined powder XRD spectra. Numbers in the parentheses indicate errors on the last significant digit.}
\begin{tabular*}{\textwidth}{@{\extracolsep{\fill}}llllll}

\hline
Sample & Constituent & Crystallographic phase (Space group) & a = b (\AA) & c(\AA) & V (\AA\textsuperscript3)                          \\ \hline
$\alpha$-Fe\textsubscript{2}O\textsubscript{3} & $\alpha$-Fe\textsubscript{2}O\textsubscript{3} & Trigonal (R-3c) &  5.045(0)           & 13.779(1)             & 303.75(3) \\ & & & & \\ 
$\alpha$-Fe\textsubscript{2}O\textsubscript{3}/ZnO & $\alpha$-Fe\textsubscript{2}O\textsubscript{3} & Trigonal (R-3c) & 5.036(0)                   & 13.754(1)             & 302.14(3)    \\
 & ZnO & Hexagonal (P63mc) &  3.250(0)                   & 5.207(2)              & 47.67(3)  \\  & & & & \\
NMS inc. $\alpha$-Fe\textsubscript{2}O\textsubscript{3}/ZnO  & $\alpha$-Fe\textsubscript{2}O\textsubscript{3} & Trigonal (R-3c) & 5.036(0)                   & 13.767(1)             & 302.93(3)   \\
  & ZnO & Hexagonal (P63mc) &  3.229(1)    & 5.227(1)              & 47.20(4)         \\  & MoS\textsubscript{2} & Hexagonal (P63/mmc) & 3.164(0)   & 12.305(0)             & 106.70(0) \\ & & & & &\\
UMS inc. $\alpha$-Fe\textsubscript{2}O\textsubscript{3}/ZnO  &   $\alpha$-Fe\textsubscript{2}O\textsubscript{3} & Trigonal (R-3c) & 5.040(0)                   & 13.764(0)             & 302.88(2)  \\
& ZnO & Hexagonal (P63mc) & 3.252(0)     & 5.203(3)   & 47.68(3)        \\  & MoS\textsubscript{2} &  Hexagonal (P63/mmc) & 3.161(1)                   & 12.300(3)             & 106.46(6)  \\\hline
\end{tabular*}
\end{table*}

\subsubsection*{Surface morphology analysis.}

The FESEM image of synthesized $\alpha$-Fe\textsubscript{2}O\textsubscript{3} nanoparticles (average size 50 to 60 nm) is shown in Fig. 2(a) which demonstrates that the nanoparticles are well dispersed with less agglomeration and their surface is satisfactorily homogeneous and non-porous. It is also clearly visible that the nanoparticles are uniform in shape and size and all the nanoparticles have trigonal crystal structure \cite{r11} (marked in the figure) which justifies the results obtained by Rietveld refined XRD patterns. Fig. 2(b) depicts the FESEM image of $\alpha$-Fe\textsubscript{2}O\textsubscript{3}/ZnO nanocomposite. It is apparent that some hexagonal shaped nanoparticles have been integrated with as-synthesized trigonal $\alpha$-Fe\textsubscript{2}O\textsubscript{3} nanoparticles which implies the successful formation of $\alpha$-Fe\textsubscript{2}O\textsubscript{3}/ZnO nanocomposite since ZnO has a hexagonal crystal structure. \cite{r19}
For getting an insight into the morphological change of MoS\textsubscript{2} due to ultrasonication, FESEM images of non-ultrasonicated and ultrasonicated MoS\textsubscript{2} have been investigated and depicted in Fig. 2(c) and 2(d), respectively. It is observed from Fig. 2(c) that non-ultrasonicated MoS\textsubscript{2} powder having comparatively rough surface morphology is composed of numerous stacked MoS\textsubscript{2} layers which cannot be detected individually. Interestingly, these stacked layers were successfully delaminated to well-dispersed ultra-thin few-layer MoS\textsubscript{2} nanosheets due to ultrasonication which is evident from Fig. 2(d).  To provide a better view, additional FESEM images of ultrasonicated MoS\textsubscript{2} sample with different magnifications have been presented in ESI Fig. S1(a) and S1(b). From Fig S1(a), using ImageJ software, the thickness of an individual MoS\textsubscript{2} nanosheet has been estimated to be in the range of 8-15 nm which corresponds to 11-21 monolayers of MoS\textsubscript{2} \cite{r10} since single-layer MoS\textsubscript{2} has a typical thickness of $\sim$ 0.7 nm.\cite{r38} Furthermore, the FESEM image of synthesized NMS incorporated $\alpha$-Fe\textsubscript{2}O\textsubscript{3}/ZnO nanocomposite is presented in Fig. 2(e). It can be seen that $\alpha$-Fe\textsubscript{2}O\textsubscript{3}/ZnO nanocomposite have been scattered over the surface of bulk MoS\textsubscript{2} powder albeit they are not well distributed and are agglomerated to a large extent. However, as demonstrated in Fig. 2(f), in UMS incorporated $\alpha$-Fe\textsubscript{2}O\textsubscript{3}/ZnO sample, the particles of $\alpha$-Fe\textsubscript{2}O\textsubscript{3}/ZnO are more uniformly distributed over the ultra-thin MoS\textsubscript{2} nanosheets as compared to the NMS incorporated sample. The small particle size of $\alpha$-Fe\textsubscript{2}O\textsubscript{3}/ZnO nanocomposite along with the nanoscopic thickness of ultrasonicated 2D MoS\textsubscript{2} sheets is anticipated to provide an increased surface area to volume ratio which is one of the essential features of an efficient photocatalyst. \cite{r6}

\begin{figure}[t!]
 \centering
 \includegraphics[width=0.45\textwidth]{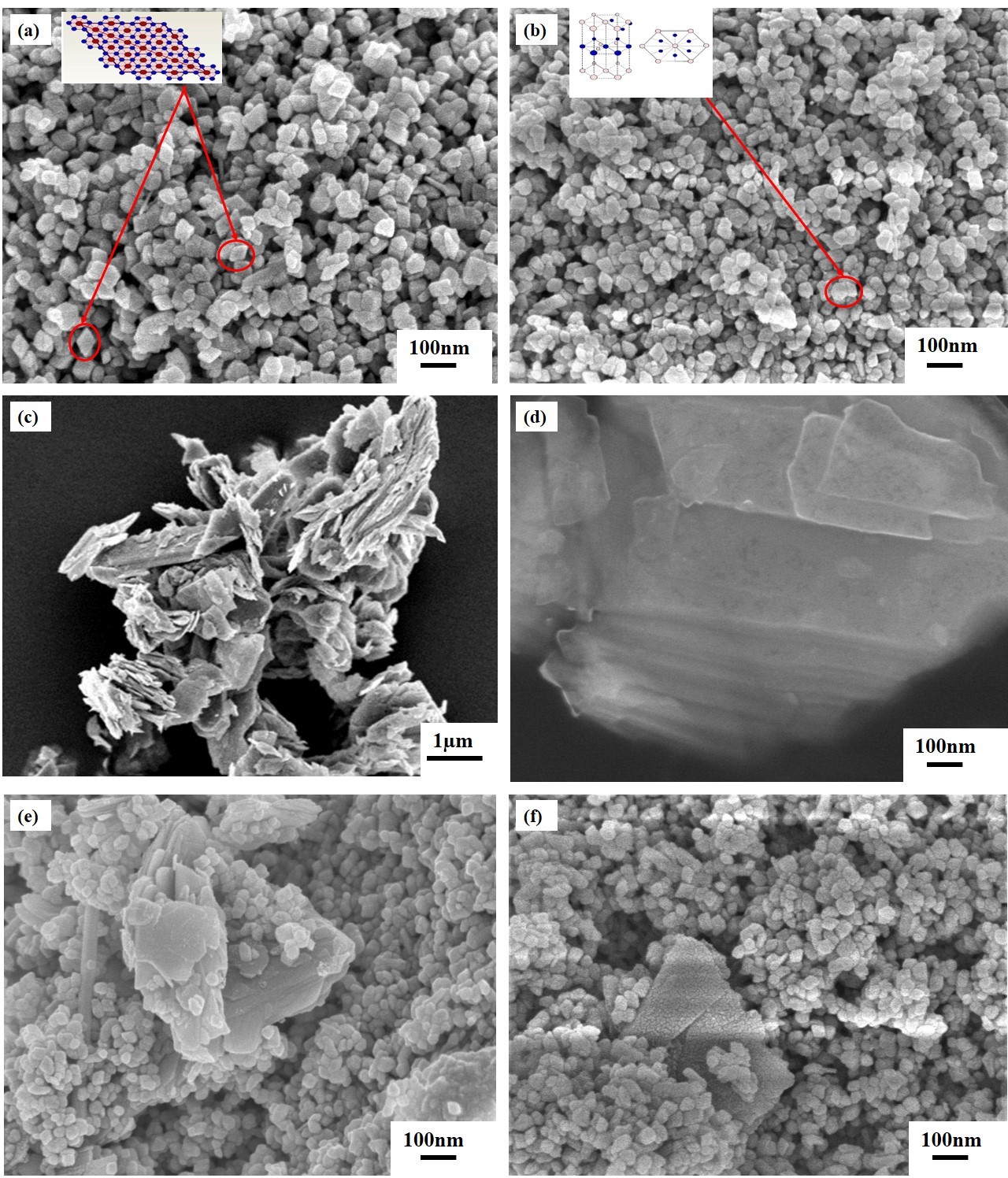}
 \caption{FESEM images of (a) $\alpha$-Fe\textsubscript{2}O\textsubscript{3} nanoparticles, (b) $\alpha$-Fe\textsubscript{2}O\textsubscript{3}/ZnO nanocomposite, (c) non-ultrasonicated bulk MoS\textsubscript{2} powder, (d) ultrasonicated few-layer MoS\textsubscript{2} nanosheets, (e) non-ultrasonicated MoS\textsubscript{2} incorporated $\alpha$-Fe\textsubscript{2}O\textsubscript{3}/ZnO and (f) ultrasonicated MoS\textsubscript{2} incorporated $\alpha$-Fe\textsubscript{2}O\textsubscript{3}/ZnO nanocomposites.}
\end{figure}

\subsection{Optical characterization}
\subsubsection*{Bandgap analysis.}

 Diffuse reflectance spectra (DRS) of all the as-synthesized samples were acquired from UV-vis spectrophotometric measurements to determine their optical bandgaps which denote the minimum energy of photons which can be absorbed by the respective material to produce e-h pairs via interband transition. It should be noted that both ZnO and $\alpha$-Fe\textsubscript{2}O\textsubscript{3} nanoparticles are widely regarded as direct bandgap materials.\cite{r28, r39} Hence, here we have applied Kubelka-Munk function and Tauc's law \cite{r13} for direct bandgap material to produce Tauc plots from the DRS spectra of corresponding materials. Fig. 3 displays the Tauc plots i.e. $[F(R)\times h\nu]^{2}$ vs. h$\nu$-curves of all as-synthesized nanomaterials which were used for optical bandgap calculation. The intercept of the tangent to the curve to the energy axis provided the optical bandgaps of corresponding materials. The direct optical bandgaps of $\alpha$-Fe\textsubscript{2}O\textsubscript{3} nanoparticles, $\alpha$-Fe\textsubscript{2}O\textsubscript{3}/ZnO, NMS incorporated $\alpha$-Fe\textsubscript{2}O\textsubscript{3}/ZnO and UMS incorporated $\alpha$-Fe\textsubscript{2}O\textsubscript{3}/ZnO nanocomposites are found to be 1.87 eV, 1.90 eV, 1.83 eV and 1.96 eV respectively (Table 2). The energy bandgaps calculated for $\alpha$-Fe\textsubscript{2}O\textsubscript{3} nanoparticles and $\alpha$-Fe\textsubscript{2}O\textsubscript{3}/ZnO nanocomposite are consistent with the values reported by Mishra and Chun (2015)\cite{r28} and Wu et al. (2012).\cite{r45} Notably, an earlier investigation \cite{r39} on ZnO nanoparticles reported their optical bandgap to be 3.37 eV. Hence, the bandgaps demonstrated by all the synthesized nanocomposites are significantly smaller compared to ZnO nanoparticles. This redshift in the bandgap indicates that prepared nanocomposites are competent for absorbing photons in the visible range of the solar spectrum more effectively which may enhance their promising potential in photocatalytic applications. The slightly enhanced optical bandgap of UMS incorporated $\alpha$-Fe\textsubscript{2}O\textsubscript{3}/ZnO nanocomposite as compared to NMS incorporated nanocomposite can be associated with the little increase in the bandgap of MoS\textsubscript{2} due to ultrasonication as reported in our previous investigation.\cite{r10} However, the observed variation among the bandgaps of the as-synthesized nanomaterials is very small. A previous investigation \cite{r6} demonstrated that the ascendancy of a certain nanomaterial as photocatalyst cannot be imputed only to its bandgap. Instead, some other determinants e.g. phase purity, surface homogeneity, superior potential to enhance charge-carriers' lifetime, favorable band structure etc. contribute to the performance of a good photocatalyst.

\begin{figure}[t!]
 \centering
 \includegraphics[width=0.45\textwidth]{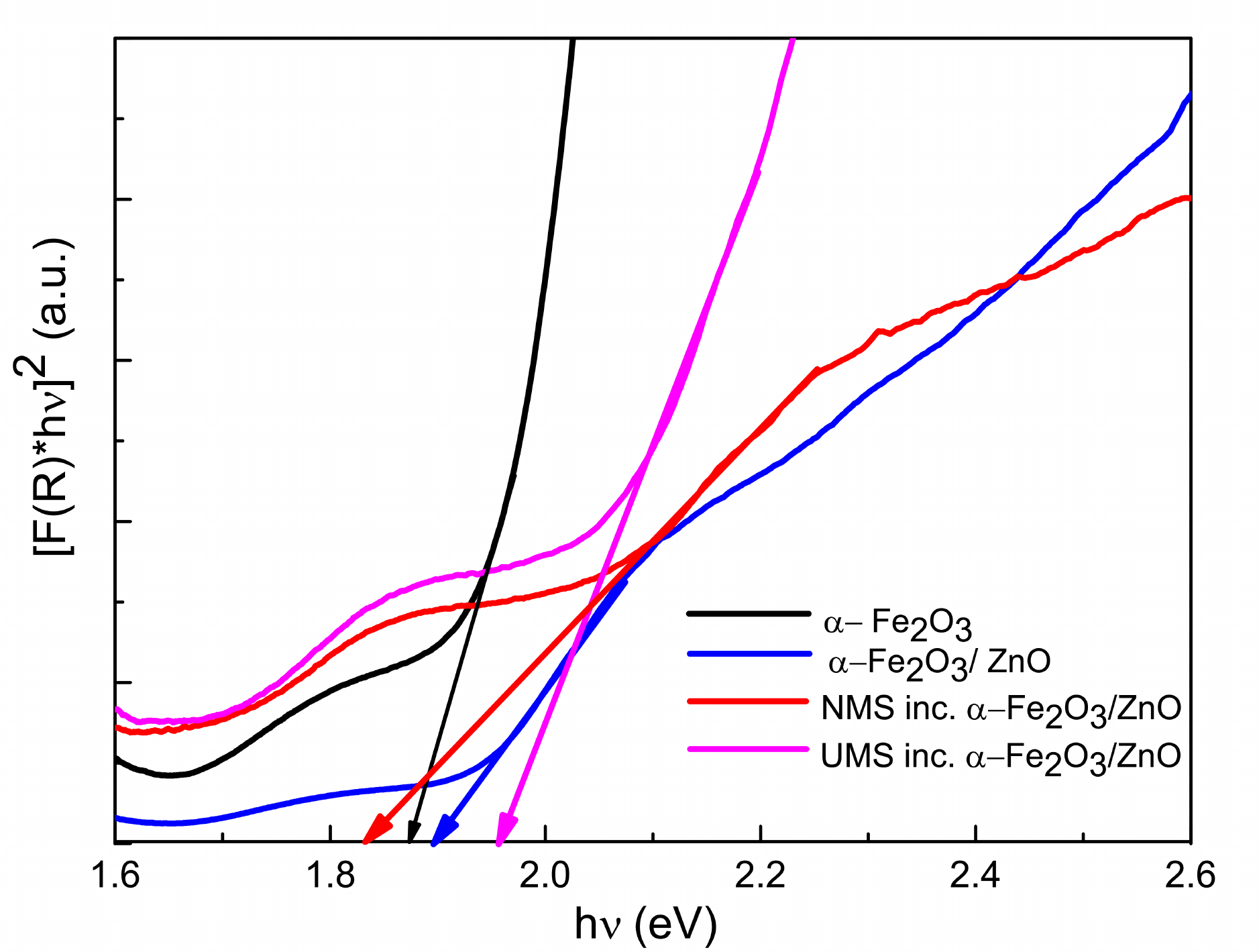}
 \caption{Bandgap estimation of synthesized $\alpha$-Fe\textsubscript{2}O\textsubscript{3} nanoparticles, $\alpha$-Fe\textsubscript{2}O\textsubscript{3}/ZnO, non-ultrasonicated MoS\textsubscript{2} incorporated $\alpha$-Fe\textsubscript{2}O\textsubscript{3}/ZnO and ultrasonicated MoS\textsubscript{2} incorporated $\alpha$-Fe\textsubscript{2}O\textsubscript{3}/ZnO nanocomposites.}
\end{figure}

\subsubsection*{Analysis of steady-state photoluminescence (PL) spectra.}

To observe e-h recombination phenomena, we have performed steady-state PL spectroscopy of all the as-synthesized samples and presented their PL spectra in ESI Fig. S2. As can be seen from the figure, the intensity of the PL signal of  $\alpha$-Fe\textsubscript{2}O\textsubscript{3} nanoparticles is substantially higher than all other synthesized nanomaterials. Among all samples, UMS incorporated $\alpha$-Fe\textsubscript{2}O\textsubscript{3}/ZnO nanocomposite has demonstrated the smallest PL peak intensity which indicates that a very small number of photoexcited e-h have been radiatively recombined in this semiconductor.\cite{r5} Therefore, we may anticipate that the photoexcited charge carriers of $\alpha$-Fe\textsubscript{2}O\textsubscript{3}/ZnO nanocomposite  were successfully separated and transported by ultra-thin MoS\textsubscript{2} nanosheets and thus the recombination process was effectively inhibited. The outcome of our investigation is also in good agreement with previous studies where 2D MoS\textsubscript{2} nanosheets have been reported to efficiently quench photoluminescence of different semiconductor photocatalysts.\cite{r47, r48} However, the radiative recombination rate of NMS incorporated $\alpha$-Fe\textsubscript{2}O\textsubscript{3}/ZnO is observed to be a bit higher than that of UMS incorporated nanocomposite, since the photoexcited e-h pairs cannot be satisfactorily segregated and transported by the stacked layers of bulk MoS\textsubscript{2} powder as compared to few-layer MoS\textsubscript{2} nanosheets.\cite{r10} 


\subsection{Photocatalytic degradation activity and stability}

The photocatalytic performance of $\alpha$-Fe\textsubscript{2}O\textsubscript{3} nanoparticles, $\alpha$-Fe\textsubscript{2}O\textsubscript{3}/ZnO, NMS incorporated $\alpha$-Fe\textsubscript{2}O\textsubscript{3}/ZnO and UMS incorporated $\alpha$-Fe\textsubscript{2}O\textsubscript{3}/ZnO nanocomposites have been investigated by a colorant decomposition test of RhB dye under solar light irradiation.\cite{r6} For comparison, titania nanoparticles had been also subject to the photodegradation analysis of RhB dye in identical experimental setup. As a representative of synthesized nanomaterials, in Fig. 4(a) we have presented RhB dye's absorbance spectra measured in dark and light conditions at 1 hour interval when UMS incorporated $\alpha$-Fe\textsubscript{2}O\textsubscript{3}/ZnO photocatalyst was employed. The intensity of absorbance peak of RhB decreases gradually with time which indicates the decomposition of RhB dye. Since RhB dye is highly resistive to decomposition, this result implies that UMS incorporated $\alpha$-Fe\textsubscript{2}O\textsubscript{3}/ZnO nanocomposite has demonstrated high photocatalytic efficiency in RhB degradation. For determining the dye degradation efficiency of the as-prepared samples both in dark conditions and after 4 hours of irradiation, the maximum intensity ratio C/C\textsubscript{0} vs. irradiation time in h is plotted in Fig. 4(b), where C and C\textsubscript{0} are obtained from the absorption spectra of RhB under stimulated solar illumination which respectively denote the highest intensity at initial time (0 hour) and at particular times (1$-$4 hours). For all the photocatalysts, we can observe that the degradation percentage is very small in dark conditions from which we can infer that the degradation of RhB molecules due to adsorption to the catalyst nanomaterials is quite negligible. Moreover, in order to evaluate the self-degradation rate of RhB, a blank test was run under 4 hours of illumination without using any photocatalyst and it was observed that only $\sim$3\% of RhB dye was degraded at the end of the experiment. On the other hand, after 4 hours of irradiation, $\alpha$-Fe\textsubscript{2}O\textsubscript{3} nanoparticles, $\alpha$-Fe\textsubscript{2}O\textsubscript{3}/ZnO, NMS incorporated $\alpha$-Fe\textsubscript{2}O\textsubscript{3}/ZnO and UMS incorporated $\alpha$-Fe\textsubscript{2}O\textsubscript{3}/ZnO nanocomposites have respectively degraded 37\%, 50\%, 53\% and 91\% of RhB dye. Under identical experimental conditions, 65\% degradation is achieved by titania nanoparticles. Since the adsorption process has very trivial contribution to the degradation of RhB as evident from the dark test, it may be inferred that the decolorization of RhB dye by irradiated catalysts has almost completely resulted from photocatalytic degradation. With a view to investigating the rate of degradation quantitatively, the data were fitted adapting a first-order model as expressed by $ln(C_{0}/C)= kt$, \cite{r24} where C\textsubscript{0} and C are the same as mentioned earlier, k is the first-order rate constant which can be considered as a basic kinetic parameter of a photocatalyst. The pseudo-first order kinetics fitting data for the photocatalytic decomposition of RhB has been shown in Fig. 4(c). The slope of the straight line provides k. Blank RhB has a very small degradation rate of $ 4.31\times10^{-3} min^{-1} $. On the other hand, the degradation rates (k) are calculated to be $ 9.84\times10^{-2} $, $ 1.66\times10^{-1}$, $ 1.94\times10^{-1}$, $ 6.25\times10^{-1}$ and  $ 2.9\times10^{-1} min^{-1} $ for $\alpha$-Fe\textsubscript{2}O\textsubscript{3}, $\alpha$-Fe\textsubscript{2}O\textsubscript{3}/ZnO, NMS incorporated $\alpha$-Fe\textsubscript{2}O\textsubscript{3}/ZnO, UMS incorporated $\alpha$-Fe\textsubscript{2}O\textsubscript{3}/ZnO and titania respectively (Table 2). Notably, UMS incorporated $\alpha$-Fe\textsubscript{2}O\textsubscript{3}/ZnO nanocomposite exhibits 277\%, 222\% and 116\% greater degradation rate compared to that of $\alpha$-Fe\textsubscript{2}O\textsubscript{3}/ZnO, NMS incorporated $\alpha$-Fe\textsubscript{2}O\textsubscript{3}/ZnO and titania respectively.

\begin{table*}
\small\addtolength{\tabcolsep}{-5pt}
\caption{Calculated bandgap energy (E\textsubscript{g}) of as-synthesized nanomaterials, photodegradation rate constant (k) and degradation(\%) of RhB dye after 4h of solar illumination in the presence of the samples}

\begin{tabular*}{1\textwidth}{@{\extracolsep{\fill}}llll}

\hline
Sample                                                                                                                 &  \begin{tabular}[c]{@{}l@{}}E\textsubscript{g}\\ in eV\end{tabular}  & \begin{tabular}[c]{@{}l@{}}Rate constant (k)\\ $\times 10^{-1} $\end{tabular} & \begin{tabular}[c]{@{}l@{}}Degradation (\%)\\ after 4 h\end{tabular} \\ \hline
$\alpha$-Fe\textsubscript{2}O\textsubscript{3} & 1.87     & 0.984                                                             & 37                                                                   \\
$\alpha$-Fe\textsubscript{2}O\textsubscript{3}/ZnO                  & 1.90     & 1.66                                                              & 50                                                                 \\
NMS inc. $\alpha$-Fe\textsubscript{2}O\textsubscript{3}/ZnO & 1.83     & 1.94                                                              & 53                                                                   \\
UMS inc. $\alpha$-Fe\textsubscript{2}O\textsubscript{3}/ZnO & 1.96     & 6.25                                                              & 91                                                       \\   
   \hline
 \end{tabular*}
\end{table*}

\begin{figure*}
 \centering
 \includegraphics[width=0.9\textwidth] {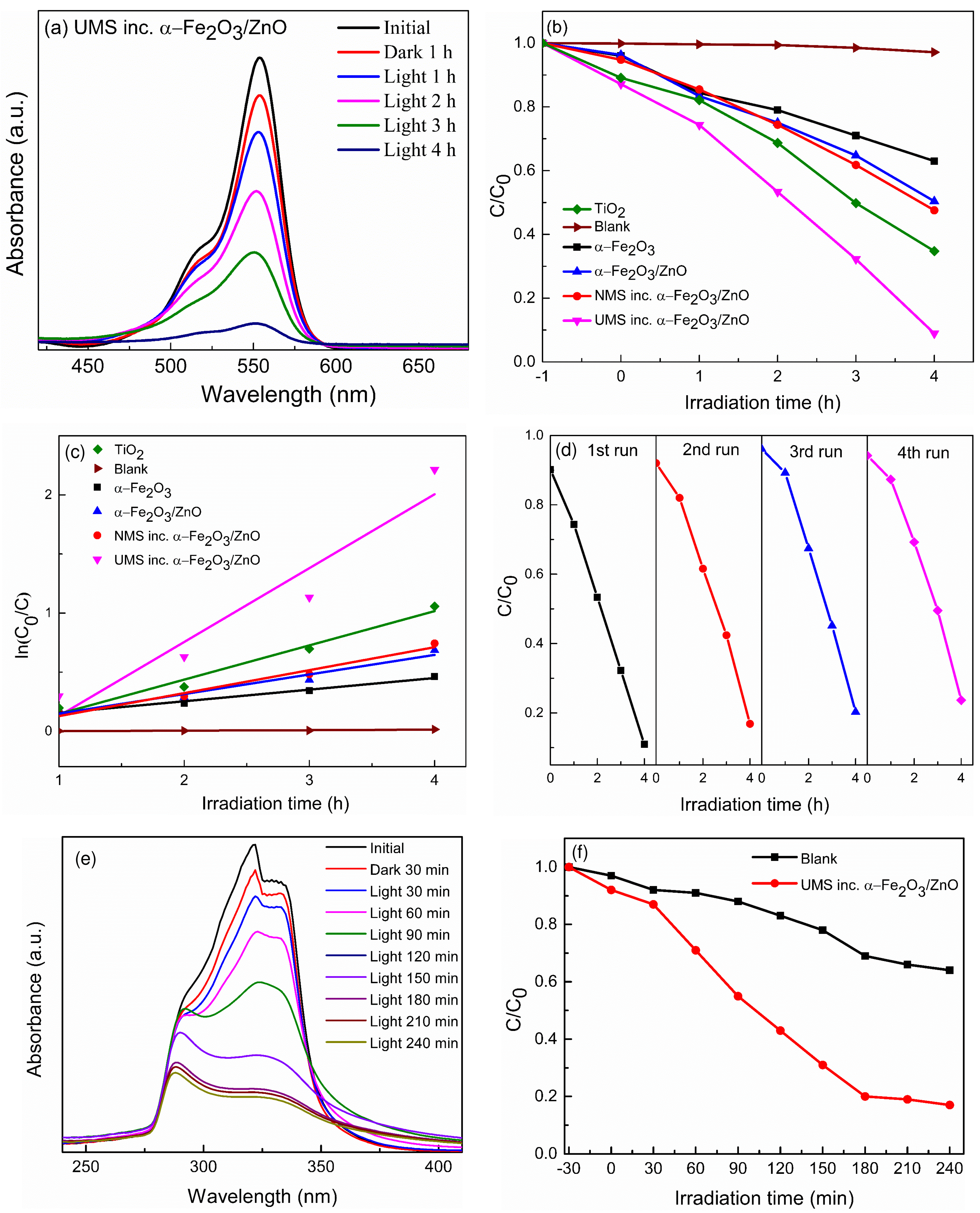}
 \caption{(a) Time-dependent absorption spectra of RhB solution after being irradiated by solar light in the presence of ultrasonicated MoS\textsubscript{2} incorporated $\alpha$-Fe\textsubscript{2}O\textsubscript{3}/ZnO nanocomposite; (b) Degradation of RhB as a function of the irradiation time for titania, blank sample, $\alpha$-Fe\textsubscript{2}O\textsubscript{3}, $\alpha$-Fe\textsubscript{2}O\textsubscript{3}/ZnO, non-ultrasonicated MoS\textsubscript{2} incorporated $\alpha$-Fe\textsubscript{2}O\textsubscript{3}/ZnO and ultrasonicated MoS\textsubscript{2} incorporated $\alpha$-Fe\textsubscript{2}O\textsubscript{3}/ZnO nanomaterials; (c) Pseudo-first order kinetics for RhB; (d) Recyclability of ultrasonicated MoS\textsubscript{2} incorporated $\alpha$-Fe\textsubscript{2}O\textsubscript{3}/ZnO for 4 successive runs of photocatalytic RhB dye degradation; (e) Time-dependent absorption spectra of ciprofloxacin solution after being irradiated by solar
light in the presence of ultrasonicated MoS\textsubscript{2} incorporated $\alpha$-Fe\textsubscript{2}O\textsubscript{3}/ZnO nanocomposite; (f) Degradation of ciprofloxacin as a function of irradiation time for blank sample and ultrasonicated MoS\textsubscript{2} incorporated $\alpha$-Fe\textsubscript{2}O\textsubscript{3}/ZnO nanocomposite. }
\end{figure*}

In addition to the superior degradation ability, the photocatalysts need to be stable and recyclable under typical photocatalytic reaction conditions for practical applications. Therefore, UMS incorporated $\alpha$-Fe\textsubscript{2}O\textsubscript{3}/ZnO nanocomposite had been subjected to a recyclability test in indentical experimental conditions and the reusability graph is presented in Fig. 4(d). As can be observed from the graph, the synthesized nanocomposite shows satisfactory stability with negligible deterioration in the photocatalytic performance after 4 successive photodegradation cycles. Finally, with a view to investigating whether any toxic by-products are produced during this degradation process, we have performed an experiment using UMS incorporated $\alpha$-Fe\textsubscript{2}O\textsubscript{3}/ZnO nanocomposite as photocatalyst under solar light illumination having wavelength ranging from 200 to 1100 nm. The result is presented in ESI Fig. S3 which shows that after degradation, no additional peaks have arisen in the absorbance spectra of RhB dye implying the absence of any foreign carbon assisted molecule or benzene radical in the solution. Since UMS incorporated $\alpha$-Fe\textsubscript{2}O\textsubscript{3}/ZnO nanocomposite has shown outstanding photocatalytic degradation ability and stability, we anticipate that it can be deemed a potential contender for practical applications such as removal of industrial pollutants from waste water, solar H\textsubscript{2} production etc.

Further, we have assessed the photocatalytic efficiency of UMS incorporated $\alpha$-Fe\textsubscript{2}O\textsubscript{3}/ZnO nanocomposite in degradation of a colorless antibiotic, ciprofloxacin under solar illumination. The absorbance spectra of ciprofloxacin determined at 30 min interval in dark and light conditions are demonstrated in Fig. 4(e). The decrement of the absorbance peaks at 280 and 320 nm \cite{cipro} with time indicates the degradation of ciprofloxacin. Notably, after 4 hours of solar light illumination, almost 83\% ciprofloxacin was photodegraded by UMS incorporated $\alpha$-Fe\textsubscript{2}O\textsubscript{3}/ZnO nanocomposite as can be seen from Fig. 4(f).

\begin{figure}[b]
 \centering
 \includegraphics[width=0.5\textwidth]{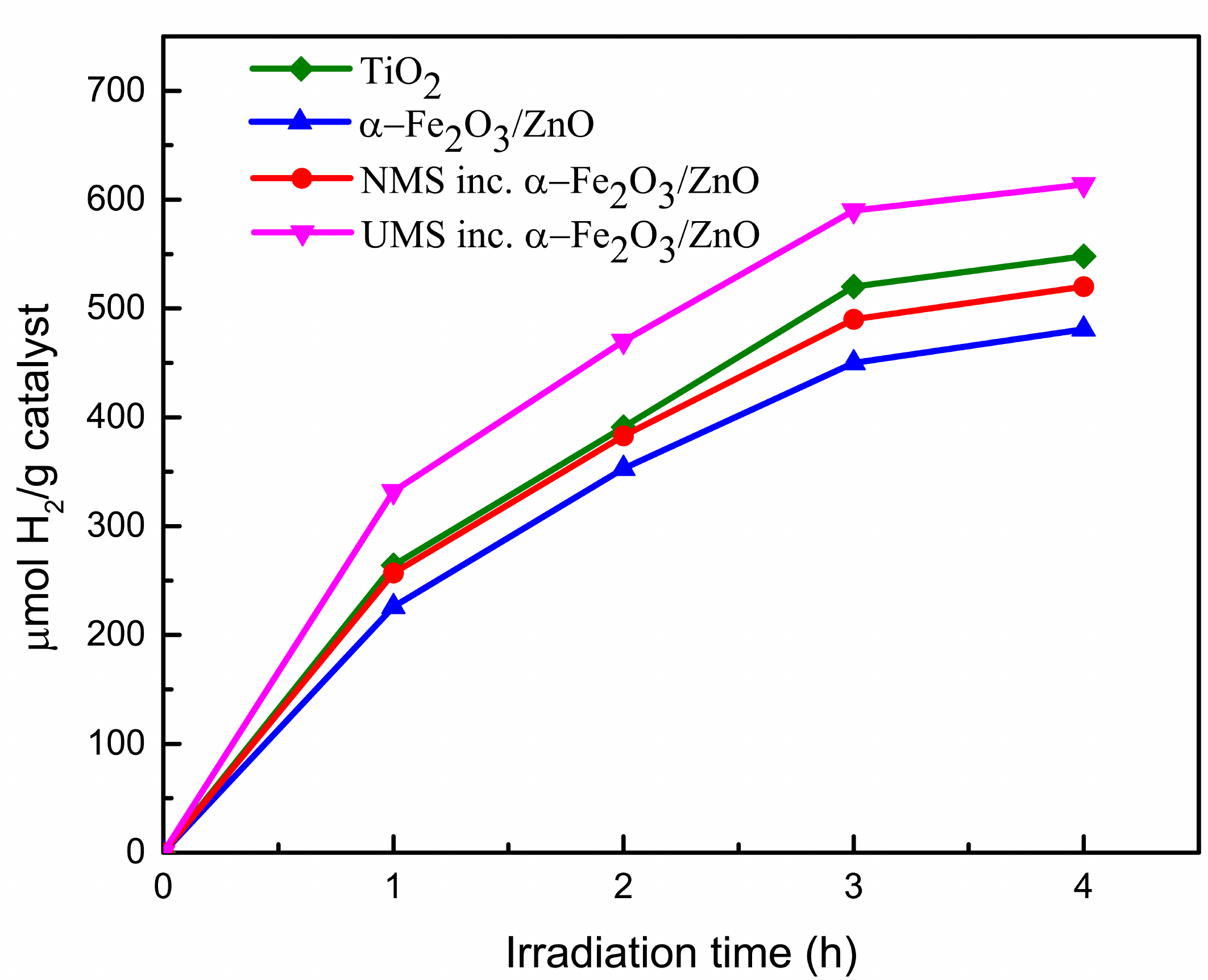}
 \caption{Volume of H\textsubscript{2} generation vs. irradiation time during water splitting for titania nanoparticles, $\alpha$-Fe\textsubscript{2}O\textsubscript{3}/ZnO, non-ultrasonicated MoS\textsubscript{2} incorporated $\alpha$-Fe\textsubscript{2}O\textsubscript{3}/ZnO  and ultrasonicated MoS\textsubscript{2} incorporated $\alpha$-Fe\textsubscript{2}O\textsubscript{3}/ZnO  nanocomposites.}
\end{figure}

\subsection{Photocatalytic hydrogen generation}

The hydrogen generation test \cite{r6, r29} has been performed using the as-synthesized nanomaterials under solar illumination. Under adapted experimental conditions, no H\textsubscript{2} evolution was observed when $\alpha$-Fe\textsubscript{2}O\textsubscript{3} nanoparticles were used as photocatalyst. Fig. 5 demonstrates the amount of H\textsubscript{2} gas (\textmu mol H\textsubscript2/g catalyst) produced by $\alpha$-Fe\textsubscript{2}O\textsubscript{3}/ZnO, NMS incorporated $\alpha$-Fe\textsubscript{2}O\textsubscript{3}/ZnO and UMS incorporated $\alpha$-Fe\textsubscript{2}O\textsubscript{3}/ZnO nanocomposites during 4 hours of solar light irradiation. For comparison, we have also performed the hydrogen generation test of titania nanoparticles and inserted the result in Fig. 5.  The results demonstrate that after 4 hours of illumination, $\alpha$-Fe\textsubscript{2}O\textsubscript{3}/ZnO nanocomposite produced $\sim$ 480 \textmu mol H\textsubscript2/g catalyst which was increased to $\sim$ 522 \textmu mol H\textsubscript2/g catalyst after the integration of non-ultrasonicated MoS\textsubscript{2}. However, the generation of H\textsubscript{2} via water splitting was further enhanced by using UMS incorporated $\alpha$-Fe\textsubscript{2}O\textsubscript{3}/ZnO nanocomposite as the photocatalyst. Notably, this nanocomposite generated $\sim$ 614 \textmu mol H\textsubscript2/g catalyst after 4 hours of solar light irradiation. The production rate is appreciably larger in comparison with all other synthesized materials as well as titania nanoparticles. The H\textsubscript{2} evolution performance of UMS incorporated $\alpha$-Fe\textsubscript{2}O\textsubscript{3}/ZnO nanocomposite is also better than that of other analogous metal oxide and TMDs based photocatalyst nanocomposites reported in literature using similar experimental conditions.\cite{r18, r44} To better understand the H\textsubscript2 generation performance level of this nanocomposite, we have provided a comparison between our work and previously reported investigations in ESI Table S2.  As shown in the Table S2, a previous investigation has demonstrated that about 190 \textmu mol H\textsubscript2 can be produced by employing MoS\textsubscript{2}/RGO nanohybrid cocatalyst under 4 hours of irradiation.\cite{r27} In another investigation on WO\textsubscript{3}/g-C\textsubscript{3}N\textsubscript{4} composite, only about 250 \textmu mol H\textsubscript{2}/g catalyst evolution has been reported after 4 hours of solar light illumination.\cite{r18} The enhancement in photocatalytic H\textsubscript{2} generation due to incorporation of ultrasonicated MoS\textsubscript{2} in $\alpha$-Fe\textsubscript{2}O\textsubscript{3}/ZnO nanocomposite can be associated with the exposed active edge sites and the accelerated charge transfer rate at the interface by few-layer MoS\textsubscript{2} nanosheets.\cite{r32} The outcome of this hydrogen production test suggests that the solar energy conversion efficiency of $\alpha$-Fe\textsubscript{2}O\textsubscript{3}/ZnO nanocomposite was improved by incorporating few-layer MoS\textsubscript{2} nanosheets and  can be further increased by integrating it with monolayer MoS\textsubscript{2} nanosheets. This is due to the fact that monolayer MoS\textsubscript{2} shows a number of unique photo-electro properties \cite{r15, r38} which may be favourable for accelerating H\textsubscript{2} evolution reaction albeit the synthesis of monolayer MoS\textsubscript{2} is quite complex and cumbersome.

\subsection{Photocatalytic mechanism}

\begin{figure}
 \centering
 \includegraphics[width=0.5\textwidth]{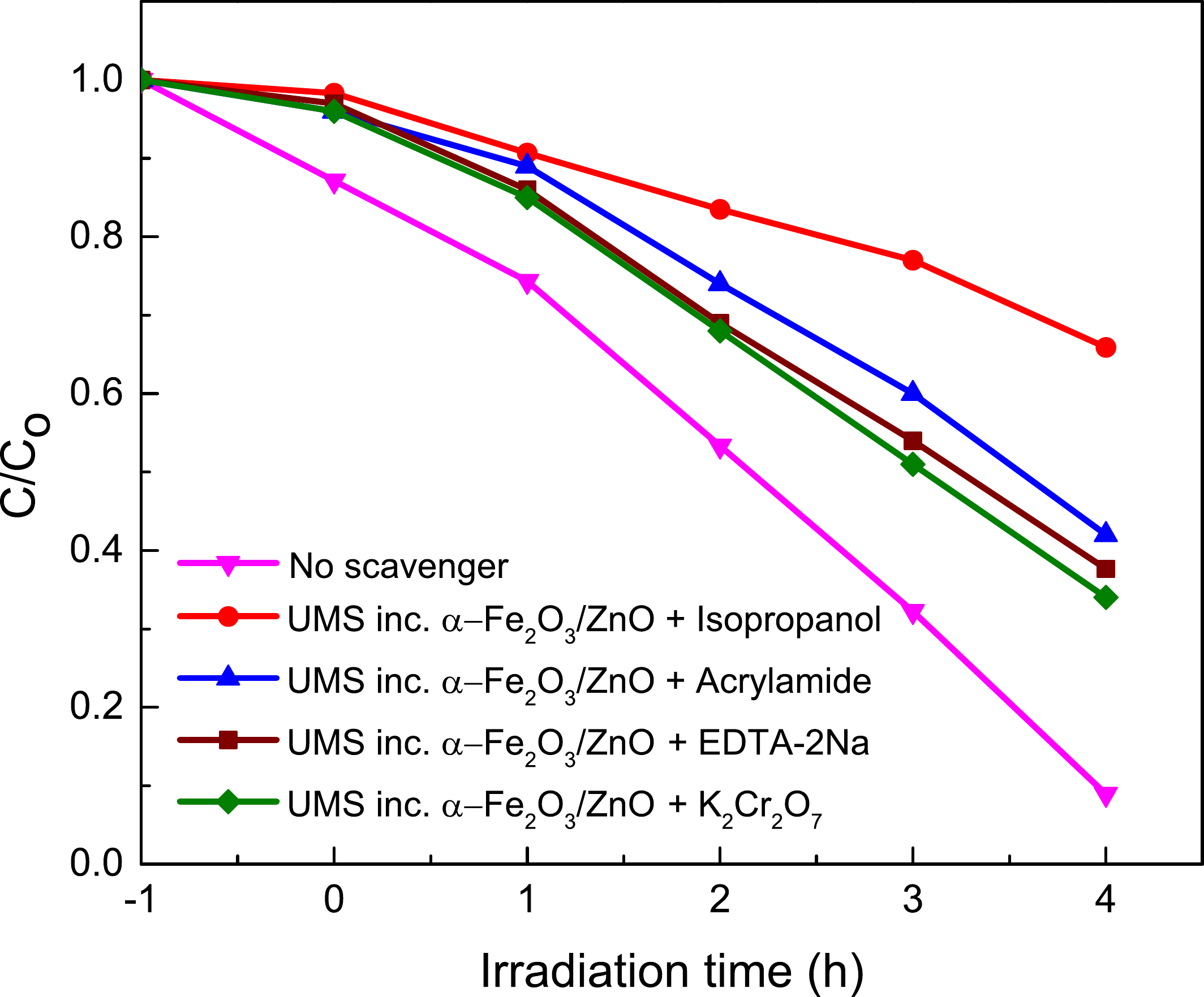}
 \caption{ Effect of the addition of different scavengers on the degradation of RhB in the presence of ultrasonicated MoS\textsubscript{2} incorporated $\alpha$-Fe\textsubscript{2}O\textsubscript{3}/ZnO nanocomposite under solar irradiation.}
\end{figure}

\begin{figure*}
 \centering
 \includegraphics[width=1\textwidth] {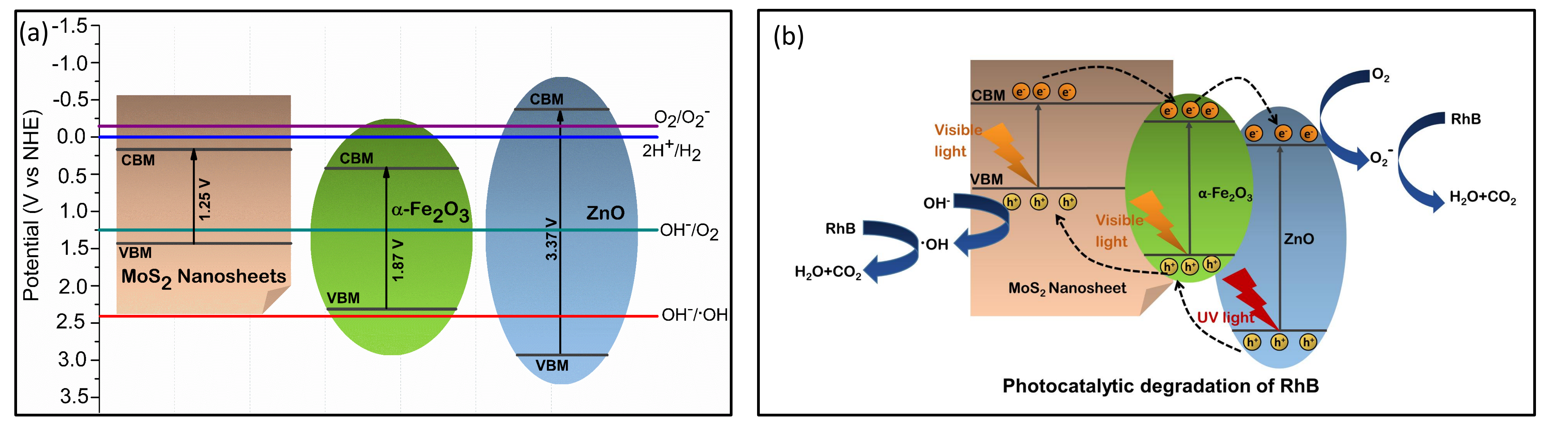}
 \caption{(a) Energy Band diagram of $\alpha$-Fe\textsubscript{2}O\textsubscript{3}, ZnO and MoS\textsubscript{2} nanosheets; (b) Schematic of proposed photodegradation mechanism of RhB using ultrasonicated MoS\textsubscript{2} incorporated $\alpha$-Fe\textsubscript{2}O\textsubscript{3}/ZnO nanocomposite as catalyst under solar illumination.}
\end{figure*}

Photocatalytic degradation of RhB dye is an electrochemical process in which photogenerated e-h from the semiconductor photocatalyst are transferred to the electrolyte (RhB solution) and then, they participate in a number of specific redox reactions to decompose the RhB solution. Two of these redox reactions are especially important and needed to be analyzed to explain the rationale behind the improved photocatalytic degradation performance of the synthesized nanocomposites.\cite{r5, r6} At first, the photoinduced holes yield hydroxyl $\cdot OH $ radical species by reacting with the $ OH^{-}$ present in the solution ionized from water (redox potential: 2.38 V vs. NHE). Eventually, $\cdot OH $ radicals oxidize and decompose RhB molecules. Meanwhile, the photogenerated electrons react with the surface-bound $ O_{2}$ along with the $ O_{2}$ in the aqueous solution (redox potential: -0.16 V vs. NHE) to produce superoxide $ O_{2}^{-} $ radicals which afterward cause degradation of RhB molecules. \\
To evaluate the contribution of different reactive species in the degradation of RhB dye, active species trapping experiments \cite{trap1, trap2} have been conducted with different scavengers in the presence of UMS incorporated $\alpha$-Fe\textsubscript{2}O\textsubscript{3}/ZnO nanocomposite and the results are presented in Fig. 6. We have employed isopropanol, acrylamide, ethylenediaminetetraacetic acid disodium salt dihydrate (EDTA-2Na) and potassium dichromate (K$_{2}$Cr$_{2}$O$_{7}$) as scavengers for trapping hydroxyl $\cdot OH $ radicals, superoxide $ O_{2}^{-} $ radicals, holes (h$^{+}$) and electrons (e$^{-}$) respectively. It is evident from Fig. 6 that the degradation percentage of RhB dye drastically reduced to 34\% from 91\% due to the addition of isopropanol into the solution of RhB dye and photocatalyst nanocomposite which is much lower as compared to other scavengers. This implies that the hydroxyl $\cdot OH $ radicals played the major role in the decomposition of RhB dye by the nanocomposite.\cite{tc1, trap1} However, the other scavengers also reduced the decomposition percentage moderately  indicating that $ O_{2}^{-}$, h$^{+}$ and e$^{-}$ were the subsidiary active species in the photocatalytic process.\cite{trap2}\\
According to the thermodynamic requirements of photocatalytic reaction, to perform degradation, the potential of photoexcited electrons and holes must exceed the redox potential of the reduction and oxidation half-reaction respectively which implies that in the case of RhB dye degradation, a photocatalyst requires to possess a valence band maxima (VBM) $>$ 2.38 V for executing the first redox reaction and conduction band minima (CBM) $<$-0.16 V to perform the second reaction \cite{r5}. Therefore, to evaluate the photodegradation performance, the CBM and VBM potentials of the prepared photocatalysts needed to be investigated and so, we have calculated the Fermi energy, CBM and VBM values of $\alpha$-Fe\textsubscript{2}O\textsubscript{3}, ZnO and MoS\textsubscript{2} nanosheets using the absolute electronegativity theory \cite{r9, r30} and tabulated the values in ESI Table S3. An energy band diagram is demonstrated in Fig. 7(a) consolidating the obtained values of band energetics along with the redox potentials of different redox half-reactions of our interest. It can be anticipated from the band structures that $\alpha$-Fe\textsubscript{2}O\textsubscript{3} nanoparticles and MoS\textsubscript{2} nanosheets cannot take part in any of the redox reactions associated with RhB dye degradation whereas the CBM and VBM positions of ZnO enable it to perform in both the reduction and oxidation half-reactions. This anticipation is strongly validated by our experimental results since the very low dye degradation efficiency (37\%) of $\alpha$-Fe\textsubscript{2}O\textsubscript{3} nanoparticles was substantially enhanced to 50\% after coupling ZnO with them.\\ 
We can propose a photocatalytic mechanism for the synthesized UMS incorporated $\alpha$-Fe\textsubscript{2}O\textsubscript{3}/ZnO nanocomposite by illumination of solar light based on the outcome of our investigation. Fig. 7(b) represents a schematic illustration of the proposed model to give a better understanding of the possible photocatalytic activities. As shown in Fig. 7(b),  $\alpha$-Fe\textsubscript{2}O\textsubscript{3} with a narrow bandgap is easily stimulated by visible light whereas ZnO absorbs only the UV portion of the solar spectrum to photogenerate electrons and holes. It can be observed from the band diagram (Fig. 7(a)) that before forming heterojunction, both the conduction band edge and Fermi level of $\alpha$-Fe\textsubscript{2}O\textsubscript{3} were lower than those of ZnO. When $\alpha$-Fe\textsubscript{2}O\textsubscript{3} comes in contact with ZnO, the energy bands of ZnO shift downward whereas the energy bands of $\alpha$-Fe\textsubscript{2}O\textsubscript{3} shift upward until their Fermi levels reach to an equilibrium and finally, a type-II heterojunction is formed at their interface where the CBM of ZnO is located between the CBM and the VBM of $\alpha$-Fe\textsubscript{2}O\textsubscript{3}.\cite{r45, type2hetero} Therefore, under solar illumination, the electrons in the VB of $\alpha$-Fe\textsubscript{2}O\textsubscript{3} will be photoexcited and transferred to the CB leaving behind holes in the VB and afterward, the photogenerated electrons will tend to move to the CB of ZnO because of the driving by the decreased potential energy. Hence, the photoinduced e-h pairs will be more effectively separated at the heterojunction which will mitigate the e-h recombination and increase the electron mobility and inter-facial charge transfer rate. Consequently, the photocatalytic performance of $\alpha$-Fe\textsubscript{2}O\textsubscript{3}/ZnO nanocomposite will be ameliorated owing to its enhanced charge separation and sunlight utilization efficiency.\\ 
Further, incorporation of ultrasonicated MoS\textsubscript{2} nanosheets increased the dye degradation capability of $\alpha$-Fe\textsubscript{2}O\textsubscript{3}/ZnO nanocomposite. According to our proposed model (Fig. 7(b)), in UMS incorporated $\alpha$-Fe\textsubscript{2}O\textsubscript{3}/ZnO nanocomposite, a heterojunction is formed between the MoS\textsubscript{2} and $\alpha$-Fe\textsubscript{2}O\textsubscript{3} which promotes the photocatalytic activity by improving e-h pair separation. During the formation of this heterojunction the electrons from $\alpha$-Fe\textsubscript{2}O\textsubscript{3} diffuse to the MoS\textsubscript{2} at the interface because of the carrier density gradient \cite{r7}. Similarly, holes from MoS\textsubscript{2} diffuse to the $\alpha$-Fe\textsubscript{2}O\textsubscript{3}. Hence, positive and negative charge regions are created in the $\alpha$-Fe\textsubscript{2}O\textsubscript{3} and MoS\textsubscript{2} sides respectively resulting in the formation of an internal electrostatic field and band bending at interface. After forming the heterojunction and aligning the Fermi levels, both the CBM and VBM of $\alpha$-Fe\textsubscript{2}O\textsubscript{3} lie under those of MoS\textsubscript{2}. Hence, under solar illumination, due to the electrostatic field, the photogenerated electrons in the CB of MoS\textsubscript{2} will transfer to that of $\alpha$-Fe\textsubscript{2}O\textsubscript{3} with higher mobility and similarly, photoexcited holes in the VB of $\alpha$-Fe\textsubscript{2}O\textsubscript{3} will easily transport to the VB of MoS\textsubscript{2}. As a result, the lifetime of the photogenerated e-h will increase which will finally result in improved photocatalytic performance. Along with modified band structure, the superiority in the photocatalytic performance of ultrasonicated MoS\textsubscript{2} incorporated nanocomposite also can be attributed to the enhanced specific surface area associated with its 2D sheet like structure and higher surface to volume ratio due to very few layer ultra-thin structure which facilitate the photocatalytic reactions by providing a good number of unsaturated surface condition sites to the RhB solution \cite{r10, r40}. The favorable effect of high surface to volume ratio on the photocatalytic activity can be further validated by analyzing the photocataytic performance of NMS incorporated $\alpha$-Fe\textsubscript{2}O\textsubscript{3}/ZnO which was not as satisfactory as the UMS incorporated $\alpha$-Fe\textsubscript{2}O\textsubscript{3}/ZnO nanocomposite. This comparatively poor performance can be ascribed to the stacked layers of non-ultrasonicated bulk MoS\textsubscript{2} powder which is not able to provide sufficient catalytically active edge sites and also cannot assist the transportation of charge carriers to the electrolyte system.\cite{r10} \\
A similar model can be suggested to delineate the underlying mechanism of the superior photocatalytic performance of ultrasonicated MoS\textsubscript{2} incorporated $\alpha$-Fe\textsubscript{2}O\textsubscript{3}/ZnO nanocomposite in solar H\textsubscript{2} evolution through water splitting. The proposed model is demonstrated in ESI Fig. S4. The photogenerated holes from the nanocomposite with sufficient potential can react with the water molecules and produce $H^{+}$ along with $ O_{2}$. Thereafter, the photoexcited electrons evolve $H_{2}$ gas from the $H^{+}$.\cite{r5, r6} Likewise the enhanced RhB degradation efficiency, it can be anticipated that the higher H\textsubscript{2} evolution rate of UMS incorporated $\alpha$-Fe\textsubscript{2}O\textsubscript{3}/ZnO nanocomposite is associated with its large specific surface area, efficient separation of photogenerated e-h pairs at the heterojunction, accelerated charge transportation and suppressed charge recombination. In addition to these, the unique structure of MoS\textsubscript{2} nanosheets may play a key role by yielding plenty of exposed edges with coordinately unsaturated S atoms as activity sites which can accept the accumulated photoinduced electrons on MoS\textsubscript{2} for reducing $H^{+}$ to $H_{2}$ gas and thus, leads to the discerned high photocatalytic H\textsubscript{2} production potential. \cite{r32}

\section{Conclusions}
We have successfully exfoliated few-layer 2D MoS\textsubscript{2} nanosheets from bulk MoS\textsubscript{2} powder via a facile ultrasonication technique and afterward, distinctively incorporated the non-ultrasonicated MoS\textsubscript{2} (NMS) and ultrasonicated MoS\textsubscript{2} (UMS) with $\alpha$-Fe\textsubscript{2}O\textsubscript{3}/ZnO nanocomposites using low temperature hydrothermal reaction technique in order to investigate their photocatalytic performance. Notably, UMS incorporated $\alpha$-Fe\textsubscript{2}O\textsubscript{3}/ZnO nanocomposite demonstrated considerably better photocatalytic performance as compared to other synthesized nanomaterials which can be associated with its enhanced potential to inhibit charge carrier recombination as revealed by photoluminescence spectra, modified band structure and higher surface to volume ratio introduced by the 2D few-layer structure of MoS\textsubscript{2} nanosheets. Besides, the abundant active edge sites with coordinately unsaturated S atoms provided by the enormous specific surface area of MoS\textsubscript{2} nanosheets also might have played significant role in boosting the photocatalytic activities of this nanocomposite. Due to the observed superior photocatalytic performance over commercially available titania nanoparticles in RhB dye and antibiotic ciprofloxacin degradation as well as in hydrogen production via disintegration of water under solar illumination, the UMS incorporated $\alpha$-Fe\textsubscript{2}O\textsubscript{3}/ZnO nanocomposite may be considered as an efficient photocatalyst. Research endeavour to incorporate MoS\textsubscript{2} monolayer with $\alpha$-Fe\textsubscript{2}O\textsubscript{3}/ZnO nanocomposite is currently underway which may demonstrate further improved performance in numerous photocatalytic applications e.g. dye degradation, solar hydrogen production via water splitting, solar water disinfection etc.   

\section*{Supplementary Information}
Additional electronic supplementary information can be found online. See DOI: \href{https://doi.org/10.1039/C9RA07526G}{10.1039/C9RA07526G}

\section*{Conflicts of interest}
There are no conflicts to declare.

\section*{Acknowledgements}
Financial support from the Ministry of Science and Technology, Government of Bangladesh (Grant No.
39.00.0000.09.14.009.2019./Phy's-25/495) and Infrastructure Development Company Limited (IDCOL), Dhaka, Bangladesh is acknowledged.

